\documentclass[12pt]{article}
\usepackage[english]{babel}
\usepackage{amsmath,amsthm}
\usepackage{amsfonts}
\usepackage{sectsty}
\sectionfont{\normalsize}
\usepackage{graphicx}
\usepackage{wrapfig}

\begin{document}

\title{\textbf{\large{Radiation of the electron bunch moving in non-regular fields}}}%
\author{Lekdar A. Gevorgian$^{(1)}$ and Valeri V. Vardanyan$^{(1,2)}$\footnote{vardanyanv@yerphi.am}}%
\date{}
\maketitle

(1) Alikhanyan National Laboratory (YerPhI),Theory Department

(2) Yerevan State University, Faculty of Physics

\begin{abstract}
The problem of spontaneous radiation of the electron bunch grazing into a charged metallic surface with randomly distributed needle shaped asperities is considered. Distances between two neighboring asperities have been described by gamma distribution. Being repealed by highly charged asperities the electrons of the bunch move along non-regular periodical trajectories in the planes parallel to the metallic surface. The spatial periods of the trajectories are random quantities which are described by the same gamma distribution. The radiation characteristics of the bunch have been obtained. It is shown that the angular distributions of the number of photons radiated from the bunch and from a single electron are the same but the frequency distribution of the bunch is being drastically changed at the hard frequency region. It is proposed to develop a new non-destructive method for investigation of the metal surface roughness. The frequency distribution of the number of photons radiated under the zero angle has been obtained. That allows to find the gain expression of the stimulated radiation.
\end{abstract}

\newpage
\section{Introduction}
\indent

The radiation of the non-relativistic electron bunch moving parallel to the diffraction grating has been investigated in the experiment of Smith and Purcell \cite{Smith}. Emitted wavelength dependence on the grating constant, bunch speed and the radiation angle is derived from the Huygens' Principle. As it is shown in \cite{Gevorgian_Vardanyan} that dependence coincides with the oscillator radiation characteristic.
\newline
\indent
An anomalously high radiation intensity has been observed during the investigations of the transition radiation of the non-relativistic electron bunch grazing into a metal surface \cite{Von Blankenhagen}. Later that anomaly has been proved in the experiment \cite{Jones}. In the experiment \cite{Harutyunian} the authors have concluded that the anomaly is conditioned by the surface roughness. The right explanation of that phenomenon has been given in \cite{Gevorgian_Korkhmazian_a}, where the radiation of the non-relativistic electron grazing into a rough metal surface has been presented as a mixture of the transition radiation and the undulator radiation conditioned by the random oscillations of the electron. In that work the random fields have been described by the Gaussian distribution. In the present work the case when the bunch electrons oscillate in the planes parallel to the metallic surface has been considered and gamma distribution has been chosen to describe the random roughness of the metal surface.
\newline
\indent
It's important for the applications to develop non-destructive methods for investigation of the metallic rough surfaces (see for example \cite{Glover}). The solution of the considered in this work problem gives an opportunity to find out information about the metallic surface roughness using the bunch radiation characteristics.

\section{Surface microundulator}
\indent

\begin{wrapfigure}{r}{0.5\textwidth} 
\vspace{-20pt}
\begin{center}
\includegraphics[scale=0.99]{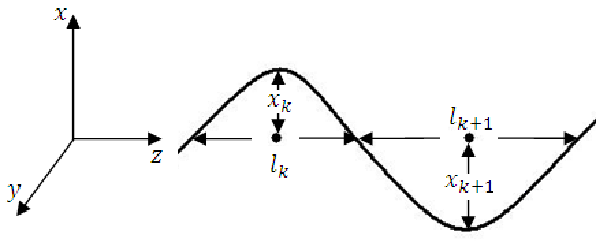}
\end{center}
\vspace{-23pt}
\caption{The path of the electron}
\vspace{-20pt}
\label{electron_path}
\end{wrapfigure}

One can assume that in our problem each electron in the bunch moves along a smooth curve like one represented on the fig. 1. On the figure by dots are represented needle shaped asperities of the metallic surface. Electrons move along $z$ direction and oscillate along $x$ direction. The distance between two asperities is $l_k$, which is also the spatial semiperiod of the electron motion.
\newline
\indent
Here we give the electron path by the following formula:

\begin{equation}
x=x_k \sin \frac{\pi}{l_k}z,
\end{equation}
\noindent
where $x_k$ is the amplitude of the trajectory.
\newline
\indent
As it is already mentioned above the trajectory should be a smooth curve for which we require the derivatives of two sins at the embroidering point be equal to each other. From that requirement it is easy to find the condition:

\begin{equation}
\beta_\perp=\frac{x_k}{l_k/2}=const,
\end{equation}
\noindent
which gives a relationship between amplitude and spatial semiperiod.
\newline
\indent
Here the following fact is very important. The bunch has a big number of electrons and at the given point of space and at the given moment of time we have photons radiated from the electrons which move along trajectories with different $l_k$. Generally speaking at the other moments of time the given electron may move along the trajectory with some other $l_k$ but at that moment with his previous $l_k$ will move some other electron. It's clear that for us it's not important from which electron the given photon has been emitted. From the mentioned feature it follows that we can assume that every electron moves along a trajectory with a constant $l_k$. This is the fact which makes possible to find out the bunch radiation characteristics.
\newline
\indent
As it is known the radiated field of the moving charge is given by the equation:

\begin{equation}
\vec{I}=\frac{1}{\beta c}\int^{L/2}_{-L/2}{[\vec{n}\vec{v_t}]}e^{i\omega t-i\vec{k}\vec{r}}dz, \beta=\frac{v}{c}.
\end{equation}
\noindent
Here $L$ is the length of the plate, $\vec{v_t}$ is the total velocity of the electron, $v$ is the electron velocity component along the $z$ axis, $\omega$ is the frequency of the radiated photon, $\vec{r}$ is a vector from the beginning of the coordinates to the electron location point, $\vec{k}$-wave vector and $\vec{n}$ is a unit vector pointing from the electron location point to the observation point.  For the $\vec{n}$ one has the following:

\begin{equation}
\vec{n}=\hat{x}\sin{\theta}\cos{\phi}+\hat{y}\sin{\theta}\sin{\phi}+\hat{z}\cos{\theta},
\end{equation}
\noindent
where $\theta$ is the polar angle and $\phi$ is the azimuth angle. Calculating the $\vec{k}\vec{r}=k \vec{n}\vec{r}=k_x x+k_z z$ and substituting that together with (4) into (3) one will get:

\begin{equation}
\vec{I}=\frac{1}{\beta c}\int^{L/2}_{-L/2}{[\vec{A}+\vec{B}\cos{(\frac{\pi}{l_k}z)}]}e^{i(\omega \frac{z}{v}-\frac{\omega}{c}z\cos{\theta}}e^{-i\frac{\omega}{c}\sin{\theta}\cos{\phi}\frac{\beta_\perp}{z}l_k\sin{(\frac{\pi}{l_k}z)}}dz.
\end{equation}
\noindent
Here the following designations have been introduced:

\begin{equation}
\vec{A}=v\sin{\theta}(\hat{x}\sin{\phi}-\hat{y}\cos{\phi}),   \vec{B}=\frac{\beta_\perp}{2}\pi v(\hat{y}\cos{\theta}-\hat{z}\sin{\theta}\sin{\phi}).
\nonumber
\end{equation}
\indent
Taking into account that in practice electrons oscillate with non large amplitudes the problem may be considered within a dipole approximation which means that in the radiation main contribution is made by the first harmonic. From the dipole approximation follows that the oscillator parameter $q=\beta_\perp \gamma\ll1$, where $\gamma=(1-\beta^2)^{-1/2}$ is the Lorentz factor, and the second exponent in (5) may be replaced by the first term of its Taylor expansion:

\begin{eqnarray}
\nonumber
\vec{I}=\frac{1}{\beta c}\int^{L/2}_{-L/2}[\vec{A}+\frac{\vec{B}}{2}e^{i\frac{\pi}{l_k}z}+\frac{\vec{B}}{2}e^{-i\frac{\pi}{l_k}z}]e^{i\frac{\omega}{v}(1-\beta\cos{\theta})z}\\
(1-\frac{\omega}{2c}\sin{\theta}\cos{\phi}\frac{\beta_\perp}{2}l_ke^{i\frac{\pi}{l_k}z}+\frac{\omega}{2c}\sin{\theta}\cos{\phi}\frac{\beta_\perp}{2}l_ke^{-i\frac{\pi}{l_k}z})dz.
\end{eqnarray}

It is well known that the angular-frequency distribution of the number of radiated photons is given by the following:

\begin{equation}
\frac{d^2N}{d\omega d\o}=\frac{e^2\omega^2}{4\pi^2c^3}\frac{1}{\hbar\omega}|\vec{I_k}|^2=\frac{\alpha\omega}{4\pi^2c^2}|\vec{I_k}|^2,
\end{equation}
\noindent
where $d\o=\sin{\theta}d\theta d{\phi}$ is the solid angle, $e$ is the electron charge, $\hbar$ is the Plank's constant and $\alpha$ is the fine structure constant.

Keeping in (6) just the terms described by the first harmonic and keeping the terms the exponents of which may become zero (the terms whose exponents may not become zero don't satisfy to the energy and momentum conservation laws), integrating the (6) and multiplying by the complex conjugate we will get

\begin{eqnarray}
\nonumber
|\vec{I_k}|^2=\vec{I_k}\vec{I_k}^*=\frac{1}{\beta^2 c^2}[\frac{\omega^2 \sin^4{\theta}\cos{\phi}^2\beta_\perp^2 v^2}{4c^2}l_k^2-\frac{\omega \sin^2{\theta}\cos{\theta}\cos^2{\phi}\beta_\perp^2 \pi v^2}{2c}l_k+\\
\frac{\beta_\perp^2\pi^2 v^2}{4}(\cos^2{\theta}+\sin^2{\theta}\sin^2{\phi})]\frac{\sin^2{(\frac{\omega}{v}(1-\beta\cos{\theta}-\frac{v}{\omega}\frac{\pi}{l_k})\frac{L}{2})}}{(\frac{\omega}{v}(1-\beta\cos{\theta}-\frac{v}{\omega}\frac{\pi}{l_k}))^2}.
\end{eqnarray}

Taking into account the fact that for big values of $L$ the last multiplier of (8) becomes a $\delta$-function one will get:

\begin{eqnarray}
\nonumber
\frac{d^2 N}{d\omega d\o}=\frac{\alpha L}{16 \pi c^2}\omega\beta_\perp^2[\frac{\omega^2 \sin^4{\theta}\cos^2{\phi}}{2c^2}l_k^2-\frac{\pi\omega \sin^2{\theta}\cos{\theta}\cos^2{\phi}}{c}l_k+\\
\frac{\pi^2}{2}(\cos^2{\theta}+\sin^2{\theta}\sin^2{\phi})]\delta(\frac{\omega}{v}(1-\beta\cos{\theta}-\frac{v}{\omega}\frac{\pi}{l_k})).
\end{eqnarray}

The last formula is the angular-frequency distribution of the number of photons radiated from an electron moving along a sinusoidal trajectory with $l_k$ spatial semiperiod. Although the number of electrons in the bunch is finite and, therefore $l_k$ is a discreet quantity, with a sufficient accuracy $l_k$ may be replaced with the $l$ continuous quantity.

\section{The angular distribution of the radiation from a single electron}
\indent

One is able to find the angular distribution of the number of radiated photons from a single electron which moves along a trajectory with a spatial semiperiod $l$ by taking into account that $d\o=\sin{\theta}d\theta d\phi=-d\cos{\theta} d\phi$ and by integrating (9) with respect to $\omega$ and $\phi$. One can notice that as a result of integration $l_k$ is being replaced with an expression which depends on $\theta$. Taking into account that relativistic electrons radiate under the small angles ($\cos{\theta}\approx1-\theta^2/2$), for the angular distribution with respect to $\Theta=\theta\gamma$ one will get:

\begin{equation}
\frac{dN}{d\Theta}=\frac{\pi^3\alpha L q^2}{4l}\Phi(\Theta),   \Phi(\Theta)=\frac{1+\Theta^4}{(1+\Theta^2)^4}\Theta.
\end{equation}
\noindent

\section{The frequency distribution of the radiation from a single electron}
\indent

It is possible to find the frequency distribution of the number of radiated photons from a single electron which moves along a trajectory with a spatial semiperiod $l$ by integrating (9) with respect to $\cos{\theta}$ and $\phi$. After simple transformations introducing the following non-dimensional frequency:

\begin{equation}
X=\omega\langle l \rangle/\pi\beta c \gamma^2,
\end{equation}
\noindent
where $\langle l \rangle$ is the mean value of the spatial semiperiods of the bunch electron trajectories (detailed information about the distribution of $l$ is given in the next paragraph), for the frequency distribution one will get:

\begin{equation}
\frac{dN}{dX}=K[1+(1-tX)^2],   K=\frac{\alpha L \pi^3q^2}{32\langle l \rangle},
\end{equation}
\noindent
where $t=l/\langle l \rangle$. If the $l$ of the electron is constant then $t=1$ and (12) coincides with the first harmonic spectrum of the undulator radiation in the vacuum \cite{Gevorgian_Korkhmazian_b}. Because the argument of the delta function (9) is zero and the cosine function is limited one has $X$ in the $(0,2)$ interval. The (12) has a parabolic shape which in this case gets a minimum value at $X=1$.

\section{Gamma distribution}
\indent

To obtain the bunch radiation characteristics one needs to choose a distribution function for the distances $l$ between two neighboring asperities. Here the gamma distribution is chosen.

\begin{equation}
f(a,t)=\frac{a^a t^{a-1} e^{-at}}{\Gamma(a)},
\end{equation}

\noindent
where $a$ characterizes the degree of non-regularity (with smaller $a$ the degree of non-regularity is getting bigger), $t=l/\langle l \rangle$, $\langle l \rangle$ is the mean value of the gamma distribution of $l$.

Gamma distribution curves for different values of a are presented on the fig. 2. The radiation characteristics of the bunch are being obtained by averaging the corresponding characteristics of the single electron radiation.

\begin{figure}[h]
\vspace{-20pt}
\begin{center}
\includegraphics[scale=0.99]{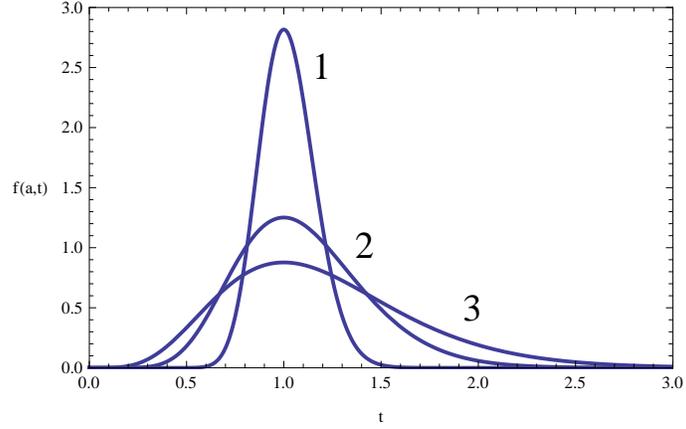}
\end{center}
\vspace{-20pt}
\caption{Gamma distribution for different values of $a$: 1) $a=50$, 2) $a=10$, 3) $a=5$}
\vspace{-5pt}
\label{gamma}
\end{figure}

It is known that the radiation line bandwidth is conditioned by the finiteness of the bunch motion and by its non-regularity as well as by the angular and energetic divergence of the bunch electrons. Usually the angular divergence is much smaller than the energetic divergence. For that case the radiation line bandwidth is determined by the following expression \cite{Gevorgian_Korkhmazian_c}:

\begin{equation}
\frac{\Delta\omega}{\omega}\approx\frac{2l}{L}+[2(\frac{\Delta \gamma}{\gamma}+\frac{\Delta l}{l})]^{1/2}.
\nonumber
\end{equation}

It is considered the case when the radiation line bandwidth is caused by the random distribution of  $l$.
\newpage
\section{The angular distribution of the bunch radiation}
\noindent

To obtain the angular distribution of the number of photons radiated from the bunch consisting of $N_b$ electrons one needs to average the (10) by gamma distribution. Since the argument of the $\delta$-function in (9) should become zero one has $t\leq2/X$. Averaging one will get:
\newline
\newline
\newline
\begin{equation}
\langle\frac{dN}{d\Theta}\rangle=N_b\frac{\pi^3 \alpha L q^2}{4\langle l \rangle} \int_0^{2/X}{\frac{f(a,t)}{t}}dt\Phi(\Theta).
\end{equation}

From the comparison of (10) and (14) it is easy to see that the angular distribution of the bunch radiation does not differ from the angular distribution of the single electron radiation which moves along a trajectory with $2l$ spatial period. That distribution is depicted on the fig. 3. Let's notice that (10) has a maximum value at $\Theta\approx0.5 (\theta\approx1/2\gamma)$.

\begin{figure}[h]
\begin{center}
\includegraphics[scale=0.97]{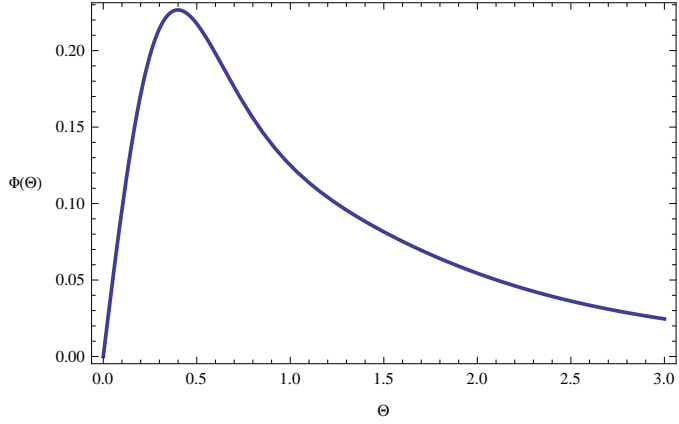}
\caption{\small{Angular distribution of the number of radiated photons}}
\label{ang_dist}
\end{center}
\end{figure}
\newpage
\section{\normalsize{The frequency distribution of the bunch radiation}}
\noindent

By averaging (12) by gamma distribution for the frequency distribution of the number of radiated photons one will get:

\begin{equation}
\langle\frac{dN}{dX}\rangle=N_b\int_0^{2/X}K[1+(1-tX)^2]f(a,t)dt.
\end{equation}

The (15) can be rewritten in the following analytic form:
\begin{eqnarray}
\nonumber
\langle\frac{dN}{dX}\rangle=KN_b[2P(a,2a/X)-2XP(a+1, 2a/X)+\\
X^2(1+1/a)P(a+2, 2a/X)],
\end{eqnarray}
\noindent
where $P(a,X)=\int_0^X{t^{a-1}e^{-t}/\Gamma(a)}dt$ is the incomplete gamma function. On the fig. 4 the numerical graphs are presented for the regular case and for the different values of $a$. As it is easy to see at the hard frequency region the spectrum drastically differs from the regular case.

\begin{figure}[h]
\begin{center}
\includegraphics[scale=0.95]{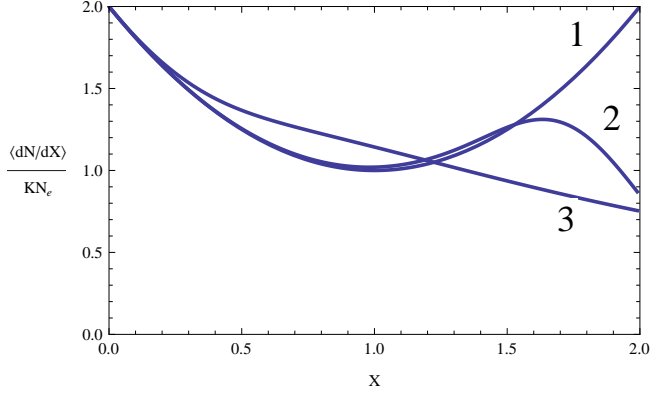}\caption{Frequency distribution of the number of radiated photons: 1) the case of undulator radiation ($a\rightarrow\infty$), 2) $a=50$, 3) $a=2$}
\label{freq_dist_1}
\end{center}
\end{figure}
\newpage
\section{\normalsize{The frequency distribution of the number of photons radiated under the zero angle}}
\noindent

A topic of a special interest is the frequency distribution of the photon number radiated under the zero angle since it uniquely defines the gain calculation of the stimulated radiation.

Putting $\theta=0$ in (9), using notation $t=l / \langle l \rangle$, multiplying (9) by the gamma distribution (13) and integrating over all possible values of $t$, for the frequency distribution of the number of photons radiated under the zero angle from the bunch one will get:

\begin{equation}
\langle\frac{d^2N}{dXd\Theta^2}\rangle=N_b\frac{\alpha L q^2 \pi^3}{8} F(a,X),   F(a,X)=\frac{a^a 2^{a-1}}{\Gamma(a)}X^{-a}e^{-2a/X}.
\end{equation}
\noindent
The obtained results for different values of $a$ are presented on the fig. 5.
\newline
\begin{figure}[h]
\begin{center}
\vspace{-20pt}
\includegraphics[scale=0.99]{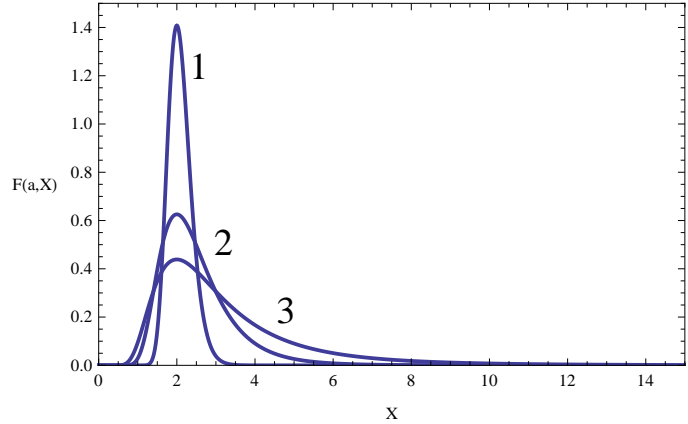}
\vspace{-10pt}
\caption{Frequency distribution of the number of photons radiated under the zero angle: 1) $a=50$, 2) $a=10$, 3) $a=5$}
\label{freq_dist_2}
\vspace{50pt}
\end{center}
\end{figure}
\newline
\newline
\newline
\newline
\newline

\section{Conclusion}
\noindent

And so the angular distribution of the bunch surface microundulator radiation coincides with the angular distribution of the single electron radiation. The frequency distribution is being drastically changed at the hard frequency region. It is proposed to develop a new non-destructive method for investigation of the metal surface roughness.
The frequency distribution of the number of photons radiated from the bunch under the zero angle has been obtained. That allows to find the gain expression of the stimulated radiation.


\begin{thebibliography}{9}


\bibitem{Smith} S.J. Smith, E.M. Purcell, Phys. Rev. \textbf{92}, 1069 (1953)
\bibitem{Gevorgian_Vardanyan} L. Gevorgian, V. Vardanyan, Il Nuovo Cimento \textbf{34 C, N. 4,} 311 (2011)
\bibitem{Von Blankenhagen} P. Von Blankenhagen, H. Boersch, D. Fritsche. H.G. Seifert, G. Sauerbrey, Physics Letters (Netherlands) \textbf{11,} 296 (1964)
\bibitem{Jones} G.E. Jones. L.S. Cram, E.A. Arakawa, Phys. Rev. \textbf{147,} 515 (1966)
\bibitem{Harutyunian} F.R. Harutyunian, R.A. Hovhannissian, A.Kh. Mkhitarian, B.O. Rostomian, Physics Letters A \textbf{43,} 107 (1973)
\bibitem{Gevorgian_Korkhmazian_a} L.A. Gevorgian, N.A. Korkhmazian, Phys. stat. sol. (b) \textbf{105,} 623 (1981)
\bibitem{Glover} J.L. Glover, C.T. Chantler, Martin D. de Jonge, Physics Letters A \textbf{373,} 1177, (2009)
\bibitem{Gevorgian_Korkhmazian_b} L.A. Gevorgyan, N.A. Korkhmazyan, Physics Letters A \textbf{74,} 453 (1979)
\bibitem{Gevorgian_Korkhmazian_c} L.A. Gevorgyan, N.A. Korkhmazyan, Sov. Phys. JETP \textbf{49,} 622, (1979)

\end{thebibliography}
\end{document}